# USING EYE TRACKER TO EVALUATE COCKPIT DESIGN -- A FLIGHT SIMULATION STUDY

Archana HEBBAR https://orcid.org/0000-0003-3772-0768[1,2], Abhay A. PASHILKAR https//orcid.org/0000-0003-2983-591X[1] and Pradipta BISWAS https//orchid.org/0000-0003-3054-6699[2]

[1]*CSIR-National Aerospace Laboratories, Bengaluru, Karnataka, India*
[2]*Indian Institute of Science, Bengaluru, Karnataka, India*
*(E-mail: archana.p@nal.res.in, apash@nal.res.in, pradipta@iisc.ac.in)*

**Abstract.** This paper investigates applications of eye tracking in transport aircraft design evaluations. Piloted simulations were conducted for a complete flight profile including take-off, cruise and landing flight scenario using the transport aircraft flight simulator at CSIR-National Aerospace Laboratories. Thirty-one simulation experiments were carried out with three pilots/engineers while recording the ocular parameters and the flight data. Simulations were repeated for high workload conditions like flying with degraded visibility and during stall. Pilot's visual scan behaviour and workload levels were analysed using ocular parameters; while comparing with the statistical deviations from the desired flight path. Conditions for fatigue were also recreated through long duration simulations and signatures for the same from the ocular parameters were assessed. Results from the study found correlation between the statistical inferences obtained from the ocular parameters with those obtained from the flight path deviations. The paper also demonstrates an evaluator's console that assists the designers or evaluators for better understanding of pilot's attentional resource allocation.

**Keywords**: Eye tracking, ocular parameters, visual scan, fatigue, IR cameras, cockpit display

## Introduction

Fifth generation aircrafts are designed to incorporate highly integrated computer systems and advanced avionics features. Design of such features require high levels of human computer interactions and hence can lead to increase in pilot's cognitive workload, if not designed intuitively. Further, the extent of usage of any display or symbology should be studied carefully to design a pilot friendly clutter free display and to ensure to provide better situational awareness. In other words, a thorough understanding of human information selection and management process is a mandatory requirement in such aircraft systems.

One of the means to understand how pilots use their visual attention resources is through remote and non-intrusive camera-based eye tracking techniques. There have been many studies conducted across the globe using eye tracking technologies. Anders (2001) recorded eye and head movements of 16 pilots in a simulator environment to assess their monitoring strategy. iView-HED+HT system from Sensomotoric instruments was used in the study wherein the eye tracking device was mounted as pilot's headband. Haslbeck et al. (2012) used a head mounted DIKABLIS eye tracking system to evaluate pilot's ability to support their manual flying skills through visual behaviour. The same eye tracker can also be used as a direct manipulation system to operate displays (Biswas & Jeevithashree, 2018)

Analysis of pilot's scan patterns can be helpful to estimate the usage of the cockpit display arrangements. This can help to achieve efficient examples of instrument scanning. Hence, eye tracking information like scan pattern can be used to improve pilot training effectiveness and efficiency. Neboshynsky (2012) conducted piloted flight experiments for age groups varying from 25-45 years and experience between 0-15+ years. The study concluded that more experienced pilots can be detected by the way they process the visual information. According to the study, experienced pilots have more refined and efficient scan patterns than novice pilots. They have a greater number of fixations and spend less time obtaining information from their displays. The researcher used three sets of non-intrusive IR cameras which were provided as input to FaceLAB 5.0. The purpose of the study was to analyze the effects of expertise and skills in navigation and target detection tasks.

Measurement of pupil dilation can be used to study pilot's cognitive load (Babu et al., 2019) (Prabhakar & Biswas, 2018) and his/her drowsiness or fatigue levels. Such an observation can be helpful as an early warning system. In a detailed study conducted by Kramer (1991), different mental workload measurement techniques using eye blinks and scan patterns were reviewed. In another study, Li et al. (2012) has evaluated the relationship between pilot's workload and their operational performance using eye tracking. Subjects used a head mounted mobile eye from Applied Science Laboratory and a head tracking device.

These research studies demonstrate that eye parameter based physiological measure is useful for the aforementioned applications. But none of these studies provide quantitative comparison of pilot's performance in terms of deviations from the desired flight path or with his/her visual scan behaviour. Pilot's performance and his/her workload can be

Using Eye Tracker to Evaluate Cockpit Design -- A Flight Simulation Study

quantitatively measured from traditional metrics using details of his/her control activity and deviations from the predefined flight profile (Hebbar & Pashilkar, 2017). Reasons for pilot's sub-optimal performance can also be gathered from the eye gaze data such as dwell time on pre-defined areas of interest, fixation and pupil diameter. So, merging both the information together can relate pilot's actions with the cockpit environment. The study presented here is an investigation to assess pilot's behaviour by means of eye gaze parameters and compare the results with his performance using statistical analysis methods.

This research work focuses on probable application areas of the eye tracking system (ETS) by means of flight simulator experiments. The experiment involves capturing the gaze coordinates and pupil diameter data from the pilot in the loop simulations conducted using CSIR-NAL's flight training simulator. Simulations were carried out for different test scenarios and flight parameters were recorded in real time. Scan pattern and dwell time on all the areas of interest (AOIs) were evaluated and compared with the flight performance data. Methodologies for calculating pilot's cognitive load and their alertness/drowsiness levels via pupil dilations are also addressed.

Another emphasis of the paper is on the hardware used in ETS. The key consideration for selection of ETS in aircraft cockpit perspective is that the system should be non-intrusive to the pilot's line of sight. While most of the system used in the above literatures uses head mounted devices, Neboshynsky (2012) uses remotely mounted IR cameras, but without illuminators. The system discussed in the present study uses 4 remotely located IR cameras and 3 near IR illuminators to provide enough light for robust gaze detection and tracking. More details on the set up and accuracy are mentioned in section 2. Funke et al. (2016) provides a detailed comparison of different eye trackers in terms of accuracy and precision.

The paper also demonstrates the development of an evaluator's dashboard that assists the designers/evaluators to understand the interactions between pilot and the aircraft interface in real-time. Designers can use this interface to reduce the demands on pilot's attentional resources (Biella et al., 2017). It can also be used to optimize different pilot vehicle interface designs. Another application of such a system would be to provide context awareness to pilots to adjust their resources when the task demand increases.

The paper begins with details of the simulation set up, selected task scenario and the defined AOIs in section 2. Section 3 presents the different case studies and analysis results. Section 4 summarizes the study and identifies the key contributions. Section 4 briefly discusses about the evaluator's dashboard. Section 6 concludes the discussions and suggests future directions.

1. Method

  1.1 Simulator Facility

High fidelity flight training simulator at CSIR-NAL for a twin engine multirole light category transport aircraft is used in this study. The simulator provides a real cockpit environment, while allowing to record flight parameters in real time. SmartEyePro ETS is installed in the cockpit as shown in Figure 1. This system measures pilot's head pose and 3 dimensional gaze direction with an accuracy of 0.5° from eye point in both azimuth and elevation and a latency < 50ms. The sampling rate of gaze stream is 40Hz. This level of accuracy is sufficient to unambiguously resolve all the defined AOIs. The system also gives additional information such as fixation, blink and pupil diameter. Latorella et al. (2010) addresses the capabilities of a similar eye tracking set up installed at NASA Langley research center.

ETS also consists of a scene camera to generate a video image of what the pilot is seeing. The scene camera video, synchronized with the data from SmartEyePro system gives a superimposed view of where the pilot is looking at. The system is ready to use after calibration of cameras and subject's gaze, which takes about 10 minutes. Care has been taken to mount the cameras and IR illuminators at locations which result in minimum obstruction to the pilot tasks.

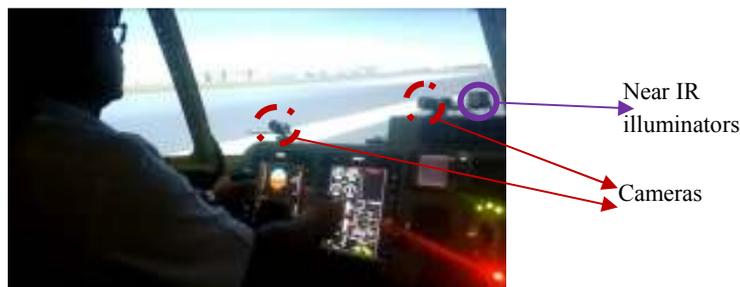

Figure 1. NAL Flight training device with cameras for eye tracking

### 1.2 Flight Profile

The task scenarios are selected such that they require significant visual scanning. Three scenarios with respect to different scan requirements are selected as follows:

i. **Nominal test case:** Pilot flies a nominal flight from take-off till landing as shown in Figure 2. The expected scan pattern during different flight phases was discussed with the pilots prior to the simulations. As per pilot's opinion, pilot tends to focus predominantly on the out of the window (OTW) visual scenery and on aircraft's attitude/altitude/airspeed in nominal flight conditions. During the throttle control, pilot's focus shall drift to torque meter. Pilot's scan behavior also differs based on the present task at hand. OTW visual scenery shall be of primary importance during lift off and landing segments. During climb phase, pilot mainly concentrates on achieving the required altitude; while maintaining attitude and airspeed. Maintaining bank angle less than 25deg is of importance during steady turns.

ii. **Degraded visibility conditions:** The visibility range is restricted to above 3000ft above mean sea level with the same task sequence as in nominal case. Pilot has to carry out manual instrument flying using display indicators in primary flight display (PFD) and Engine indication and crew alert system (EICAS) panels.

iii. **Stall case:** The stall case flight segments are similar to nominal test condition except for the following changes in the level 1 and 2 segments:
   - Level segment 1: After initial climb, pilot continues to climb till 6000ft indicated on altimeter. Pilot has to maintain this altitude and speed of 130knots.

   - Level segment 2: After completing the roll segment 1, pilot needs to commence stall by dropping speed. He/she has to monitor the rate of climb/descent (ROC) and stall warning on EICAS till stall and recovery. Once the aircraft is brought back from stall, pilot continues to descend to 4000ft at 130kts speed. He/she then levels out at 4000ft and continues with the circuit.

Pilot's scan of angle of attack (AOA) and ROC parameters shall be more predominant during stall phase. As per pilot's opinion, pilot shall monitor torque indicator more frequently in the stall phase because he/she will be using throttle to recover from stall. Figure 3 shows the flight profile for one nominal case.

### 1.3 Participants

Thirty-one simulations were carried out - one set with an expert pilot and 2 sets with test engineers. The pilot was a highly experienced military pilot who had experience in flying twenty-five different types of aircraft and had over 3400 flight hours of flying. Both test engineers were acquainted in flying this aircraft in simulation environments.

### 1.4 Procedure

All simulations were conducted in the morning (9am -12pm) and with same ambient lighting conditions. Data gathered includes all the required flight parameters such as aircraft position, orientation, speed; control column deflections (throttle, elevator, aileron, rudder, toe brakes, trim switches) and ocular parameters (head pose, three-dimensional gaze direction, pupil diameter). All data were synchronized with the start of the simulation. Cockpit audios and videos were also recorded to re-establish the operational context during analysis.

Twenty-two simulations were carried out under nominal conditions (10-expert pilot, 8-test engineer1 and 4 test engineer 2). Out of these, three cases could not be analyzed because of data loss due to reduced brightness levels and due to inaccurate time stamping between aircraft and eye tracking data. Nine more simulations were carried out with expert pilot for the high workload scenarios mentioned above. Seven are the stall cases and two are with degraded visibility conditions.

Using Eye Tracker to Evaluate Cockpit Design -- A Flight Simulation Study

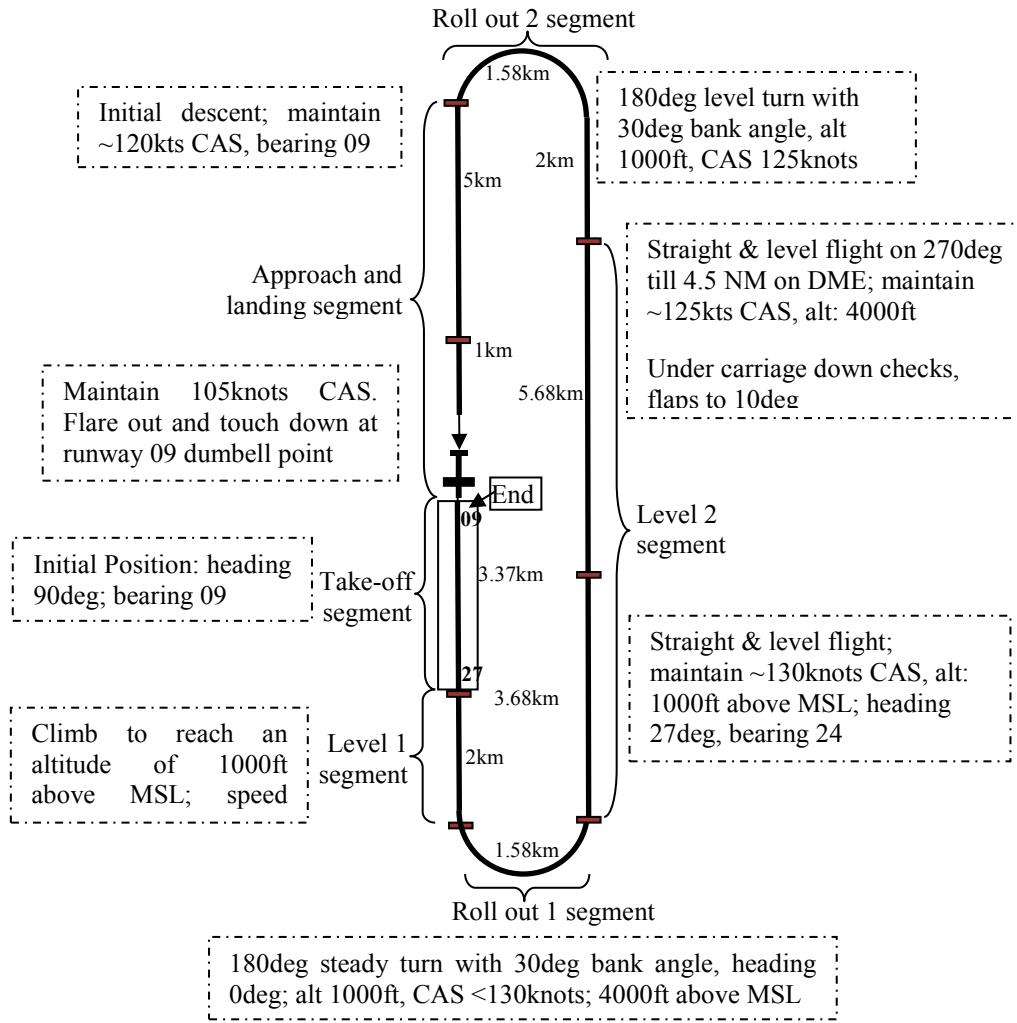

Figure 2. Test Scenario

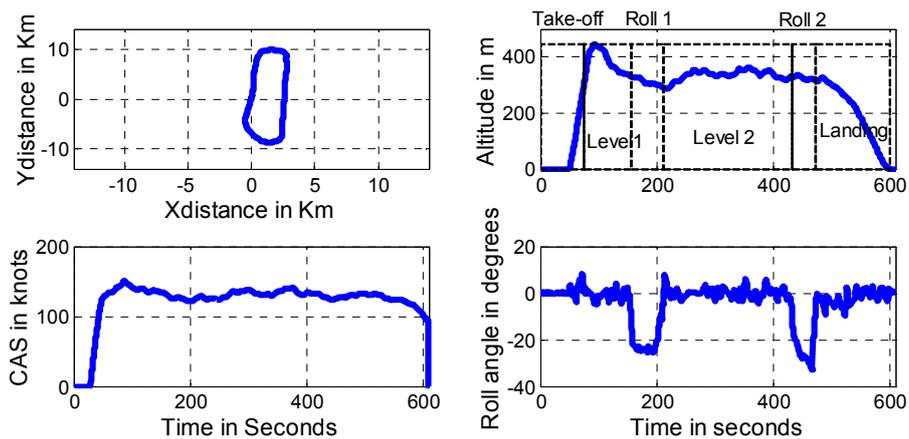

Figure 3. Flight profile

## 1.5 Areas of interest

Pilot's field of view was divided into different AOIs where pilot's monitoring is required. Primary panels that were chosen are PFD, EICAS, radio tuning unit (RTU), Autopilot panel, Integrated Standby Instrument System (ISIS), landing gear switch and OTW screen (Figure 4).

Using Eye Tracker to Evaluate Cockpit Design -- A Flight Simulation Study

PFD and EICAS in main instrument panel (MIP) were further divided into more detailed levels based on the different display scales.
1. PFD:
   A1-Airspeed Indicator; A2-Altitude Indicator; A3-Attitude scale; A4-AOA indicator; A5-Heading compass; A6-ROC indicator; A7-CRS & Ground speed indicators.

2. EICAS:
   A1-Torque indicator; A2-ITT, Ng/Np and Oil temperature indicators; A3-Master warning panel; A4-Flaps and landing gear indicators; A5-Elevator and rudder trim indicators; A6-Electrical/Fuel page.

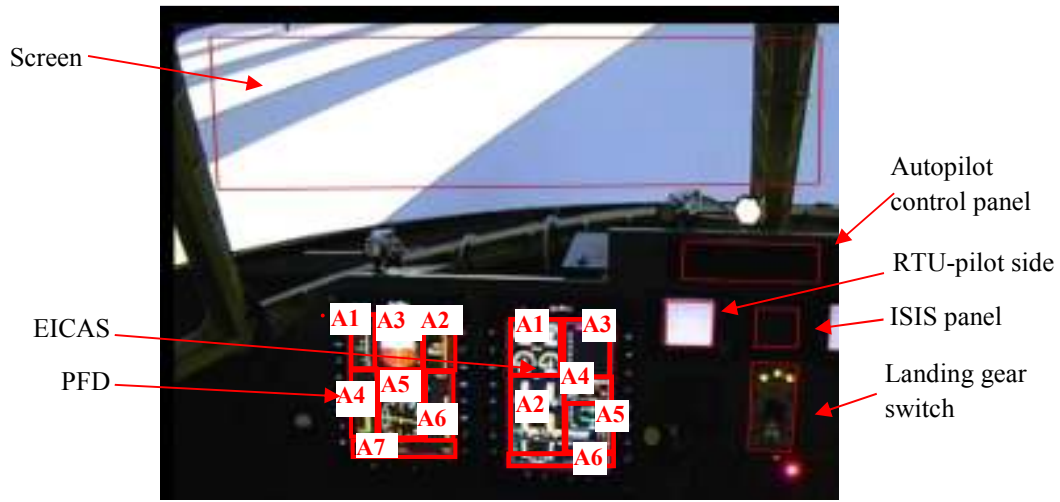

Figure 4. Selected AOIs

## 2. Analysis results

This section describes the different ocular parameter-based metrics used for analysis. Analysis results based on the potential areas of application for ETS are later discussed in detail. Outliers in the data are removed using outer fencing. Anderson-Darling test was conducted to check that the data is normally distributed. Accordingly, one way ANOVA and TukeyHSD post-hoc pairwise test was used to analyze the significance of difference between the test conditions.

Following metrics were used for gaze analysis:
1) Metrics for determining pilot's scan behavior
   a) Mean fixation map - Fixation occurs when the gaze rests for 80-100ms on a predefined area. It determines a person's focus and level of attention. The intersection points of X and Y gaze direction vectors in 3-dimensional space within the predefined AOIs is mapped onto 20X20 pixel square and is drawn on top of the scene camera image.

   b) Percentage dwell time (PDT) - Dwell time is represented by one visit in an AOI. It is taken as the sum of all fixations and saccades that hit the AOI. Fixation is the state when eyes remain still over a period of time. Saccades are rapid motion of the eye from one fixation to another. PDT time is computed as the dwell time with respect to total simulation time (Vansteenkiste et al., 2015). PDT provides information regarding the importance given by the pilot to a particular symbology/display.

2) Workload metrics - Fatigue or drowsiness is measured to be maximum if eye is at least 70% closed. Power spectral density (PSD) of normalized pupil dilations is used to measure pilot workload. Details on these metrics is discussed in section 3.3.

Proven performance-based workload analysis metrics such as root mean square error (RMSE), Pilot Inceptor workload (PIW) (Hebbar & Pashilkar, 2016) and Power frequency (Lampton & Klyde, 2012) were used to establish relation between pilot's performance and his/her physiological measures. Statistical methods such as analysis of variance (ANOVA) and paired t-test were used to establish the significance of variation amongst samples.

### 2.1  Analysis of pilot's visual scan behaviour

# Using Eye Tracker to Evaluate Cockpit Design -- A Flight Simulation Study

Analysis of pilot's scan behavior is helpful to understand how pilots handle their visual resources, what is their situational awareness or how they interact with the cockpit instruments. Firstly, scan pattern of all the 19 nominal cases is compared and correlation amongst different AOIs are observed. Figure 5 shows the PDT for all nominal cases against respective primary AOIs. The period where pilot's dwell is not in any of the defined AOI is computed as 'OTH'. This region means that either pilot's attention is in undefined areas where there is no useful information to monitor or he is moving between AOIs.

It is observed that pilot's scan in all cases is predominantly focused on OTW Screen and PFD. These are the displays required for basic flying operations. EICAS display is used for monitoring torque and engine parameters. One way ANOVA on PDT of OTW, PFD and EICAS shows that there is statistically significant difference amongst the usage of the three displays ($F(2,54)=68.76$, $p <0.001$, $\eta^2 =0.7181$ ). Here, 'p' is the probability of false positive. A TukeyHSD post-hoc test after ANOVA reveals that dwell time is not significantly different between OTW and PFD ($p = 0.8998$). However, pilot's usage of EICAS display is significantly less (OTW: 28.0+8.4%, $p < 0.01$; PFD: 26.8+11.3%, $p < 0.01$).

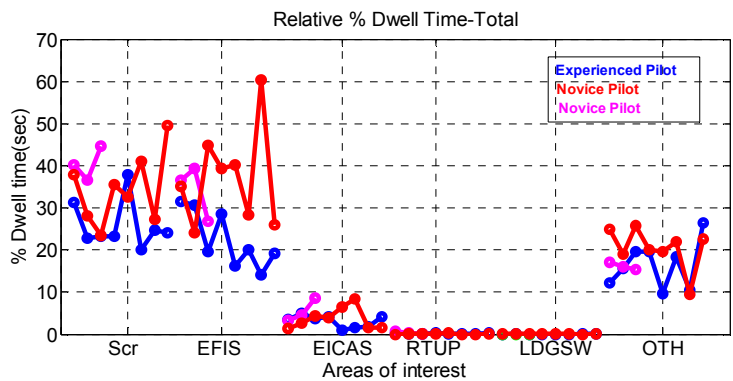

Figure 5. PDT of all nominal cases

Another observation is that experienced pilot shows lower dwell durations for OTW ($F(1,14)=5.52$, $p < 0.05$, $\eta^2 = 0.2828$ ) and EFIS display ($F(1,14)=9.37$, $p < 0.05$, $\eta^2 =0.4008$ ) when compared to novice pilot. This is because test engineers (termed as novice pilots) take longer time to decode and process information from the displays than experienced pilots. Nevertheless, this observation needs to be concluded with more simulations.

The differences in visual scan patterns for the three different task conditions are further observed using mean fixation maps and dwell time. Figure 6 shows the mean fixations for all the predefined display positions computed over the entire simulation duration. The results shown here are from the simulations conducted by the experienced pilot. Red denotes the areas that are scanned most of the time. Blue denotes the least scanned areas. It can be seen that the scan pattern is mostly focused to upper part of PFD (attitude scale, altitude and airspeed displays), torque display in EICAS and OTW screen in nominal case. In case of stall, pilot scans AOA, ROC and torque indicators predominantly for information on speed. It can be observed from Figure 6(c) that distribution of pilot's scan on the screen is very less when the visibility is low. Pilot monitors more of the cockpit displays in the low visibility scenarios.

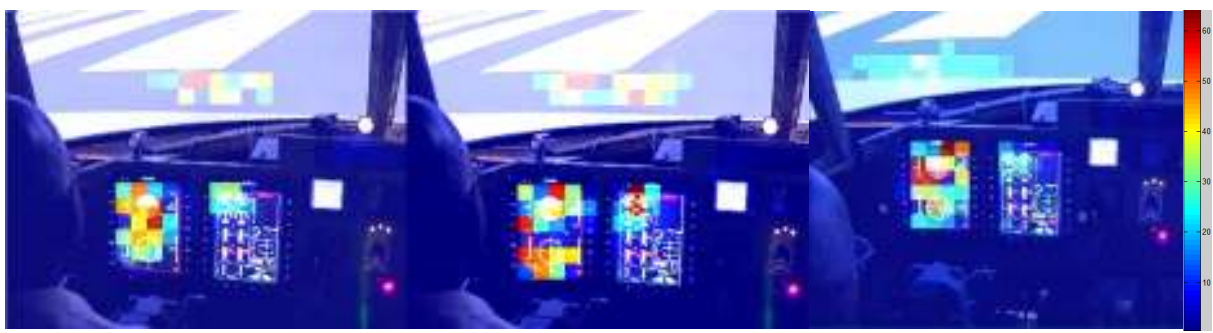

(a) Nominal case      (b) Stall case      (c) Low visibility case

Figure 6. Mean Fixation map

Using Eye Tracker to Evaluate Cockpit Design -- A Flight Simulation Study

PDT computation of nominal case (Figure 7) shows that pilot has mainly concentrated on the OTW screen (30.6% of total time). Other instrument displays that are observed often are attitude scale (10.8%), altitude scale (7.75%), heading scale (4.43%), velocity scale (3.87%) and torque meter (2.97%).

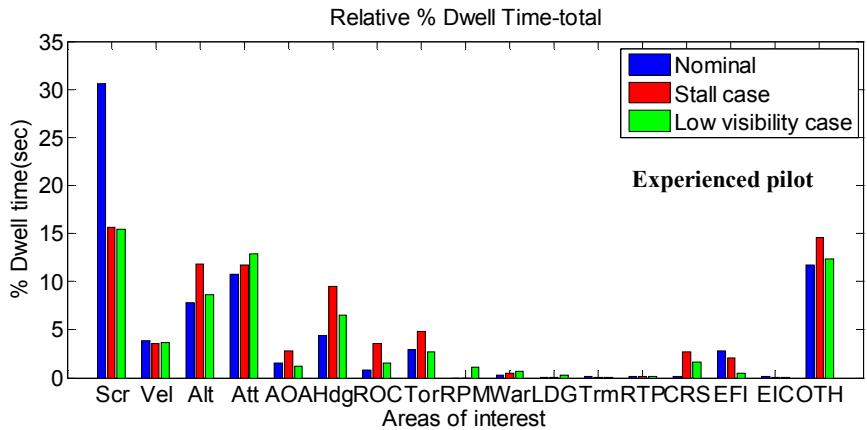

Figure 7. PDT for all three cases

In case of stall, pilot observed the OTW visuals for only 15.7% of total time. Other instruments that were observed more often are attitude scale (11.7%), altitude scale (11.8%), heading scale (9.5%), velocity scale (3.55%), torque meter (4.86%), AOA scale (2.8%) and ROC scale (3.54%), warning display (0.5%).

Hence it is clear from the above observations that we can corroborate pilot's actions with their monitoring behavior using ocular metrics such as mean fixation maps and PDT.

## 2.2 Usefulness of a display in different phases of flight

Randomly chosen 2 nominal, 2 stall and 2 cases with low visibility from the simulations conducted by experienced pilot (P1) and 2 simulation data each from both the novice pilots (P2 and P3) are used for comparing the flight path accuracy. Comparisons are done to indicate whether the errors in the flight path could be accounted to be due to lack of monitoring of the relevant parameter on the display. Straight and level flight segment 1 is discussed here.

It is desirable in this phase of flight for the pilot to maintain an altitude of 4000ft and speed of 130knots, while maintaining a constant heading. Path deviation is computed using RMSE (Figure 8). Figure 9 gives details of PDT in all major AOIs during this segment.

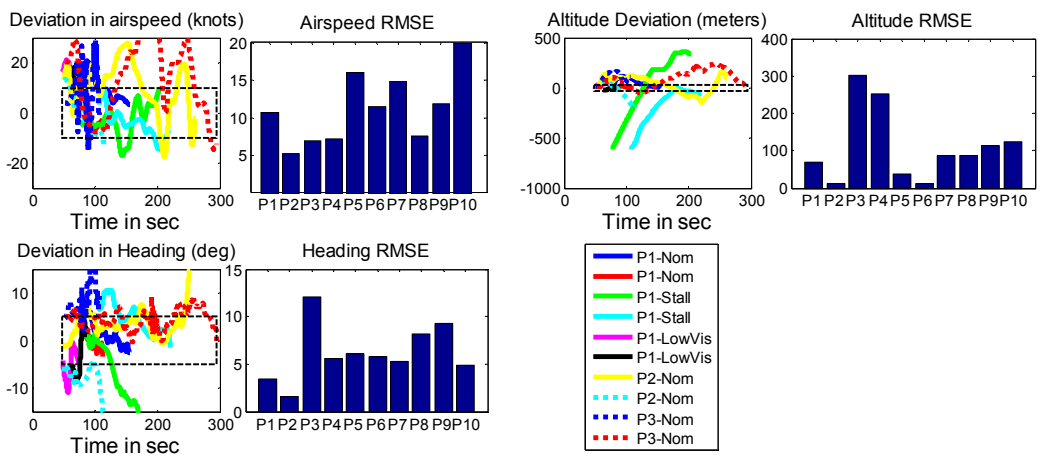

Figure 8. Error in first level segment

T-test with 0.05 level of significance (p) is computed to compare nominal and stall groups with low visibility group. The results (Figure 8) on airspeed states that the group means are significantly different (T-value = 3.5475, p <0.05, d = 2.5085), where p is the calculated probability and d is the Cohen's effect size. Deviation is more in low visibility condition (P5). Accordingly, monitoring pattern on the instruments in this case shows that altitude and attitude scales

Using Eye Tracker to Evaluate Cockpit Design -- A Flight Simulation Study

are predominantly scanned. PDT in velocity scale is less (2.5%) when compared to other AOIs. It should be noted here that visibility is near zero at this altitude and pilot's only cues for maintaining desired tolerances are from the head down displays. Hence it can be inferred that pilot has not monitored airspeed consistently during this period.

RMSE due to simulations by experienced pilot is also compared with respect to novice pilots to understand the difference in monitoring behaviors. T-test (N=18) shows that the errors due to novice pilots for a nominal scenario shows no significant difference with respect to altitude (T-value = 2.3959, p <0.05, d = 1.1979) and airspeed (T-value = 2.3838, p<0.05, d = 1.1919), which are the primary flight parameters to be controlled in the longitudinal direction. However, the errors due to novice pilots are larger for heading (T-value = 3.6759, p<0.05, d = 1.8380), which is the primary monitoring parameter for lateral axis control. This shows that novice pilots have poorer control when performing a dual or concurrent task, that is, when controlling both longitudinal and lateral axes together. This observation can be correlated with the percentage dwell durations shown in Figure 9. Both the novice pilots have predominantly used airspeed and attitude displays as compared to heading display. Attitude display is used to maintain level flight. However, experienced pilot's dwell time is distributed between airspeed, altitude, attitude and torque displays.

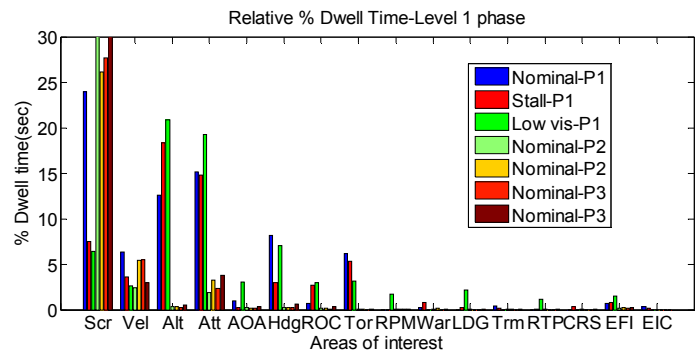

Figure 9. PDT in first level segment

Likewise, pilot's desired action is to climb to an altitude of 6000ft while maintaining wings level in stall case. It is clear in the scan pattern that pilot's gaze is concentrated maximum on the altitude scale (18.4%). As he is still climbing, altitude error is high. Due to these actions, pilot (P3) has not monitored the heading scale (3%) adequately. Hence RMSE in heading for P3 is high.

## 2.3 Pilot workload

Literature reveals (Petkar et al., 2009), (Biswas & Langdon, 2015) (Marshall, 2002) that analysis of pupil dilation and blink is a measure of pilot workload. Pupil dilates with increase in cognitive workload (Xu et al., 2011). Range of pupil dilation varies between 2mm to 8 mm (Spector, 1990) (Kret & Sjak-Shie, 2019). In order to study the effect of pupil dilation on cognitive load, normalized pupil diameter of the each of the test cases are plotted in Figure 10(a) (P1-Experienced pilot; P2, P3-novice pilot). It can be seen that the average pupil diameter is 0.92 (mean: P1-Nominal=0.92, P1-Stall=0.93, P1-low visibility=0.87; P2: 0.9418, P3: 0.9593). Novice pilots show a relatively larger pupil dilation. More metrics are further investigated to quantify any signature of workload.

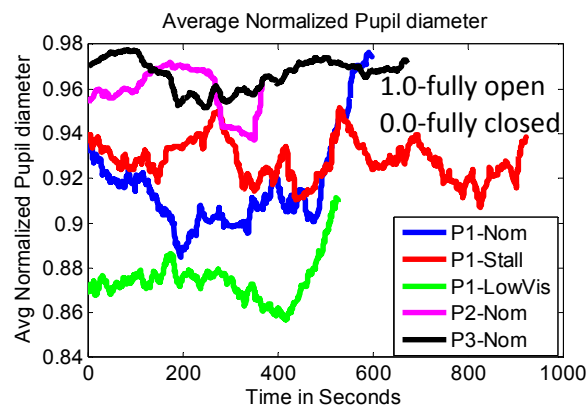

Figure 10. Normalized pupil diameter

Using Eye Tracker to Evaluate Cockpit Design -- A Flight Simulation Study

In a study, Reiner and Gelfeld (2014) have estimated mental workload using two metrics based on power spectrum of pupil fluctuations: 1. Ratio between low frequency (LF from 0.05- 0.15Hz) and high frequency (HF from 0.15-0.45Hz) components of pupil fluctuations (LF/HF ratio) and 2. High frequency components retrieved from the power spectrum of pupil fluctuations. The increase in mental activity is correlated with an increase in LF/HF ratio and a decrease in HF value.

Power spectral density (PSD) was computed using Welch periodogram method in Matlab. Mean power in the different frequency bands mentioned above and LH/HF ratio is computed (Table 1). It can be observed from the table that novice pilot's perceived mental workload is higher than that of experienced pilot.

Table 1. Mean PSD values for different frequency bands [Amp$^2$/Hz]

| Test case | LF band | HF band | LF/HF ratio |
|---|---|---|---|
| Pilot 1-Nominal case (Experienced pilot) | 0.012435 | 0.007911 | 1.571761 |
| Pilot 2-Nominal case (Novice pilot) | 0.014166 | 0.004847 | 2.922806 |
| Pilot 3-Nominal case (Novice pilot) | 0.002078 | 0.001196 | 1.737664 |
| Pilot 1 - Stall | 0.009006 | 0.005037 | 1.787988 |
| Pilot 1 - Degraded visibility simulation | 0.008478 | 0.005566 | 1.523068 |

It can also be seen that in case of increasing task difficulty (from pilot 1 alone), the stall case has the maximum LF/HF ratio and lowest HF value. Hence in the three simulations analyzed here, stall case shows an increased trend of mental activity by the pilot. This in turn indicates a higher induced cognitive load to the pilot. This correlates the pilot's feedback that both physical activities on the controls and monitoring requirement is more in the stall case.

However, it may be noted that pupil dilation also depends on many other situation-based parameters such as adjustments to different lighting conditions, pilot's physical state, fatigue and boredom. Hence pupil dilation alone cannot be considered as a measure to detect cognitive workload. It has to be corroborated with other physiological methods such as EEG or heart rate variability.

## 2.4 Monitor pilot's alertness and fatigue levels

Monitoring pilot's fatigue levels can be helpful as an early warning system. Long duration flying induces pilot fatigue. Scenarios were created in the simulator for continuous simulations for long durations. Some of the results are discussed here.

Three cases with increasing demand are simulated with a novice pilot. The pilot conducting this study was an engineer with no prior experience in flying. Before doing the actual simulations, he had flown for more than one hour to get acquainted with the controls and flying techniques. Four simulations with varying degrees of difficulty are conducted. One is the nominal condition as discussed earlier. Second is the low visibility conditions and the third is the low visibility with turbulence. At last, one nominal case is repeated to examine whether the pilot is fatigued. All the simulations are conducted continuously without any break. During the last simulation under nominal simulation conditions, the engineer had admitted that he was feeling fatigued.

Pupil diameter was measured using ETS. Figure 11(a) shows the pupil diameter, smoothened using sliding average method. Pupil diameter generally decreases and its fluctuations increase (i.e., more changes in pupil size) during drowsiness (Soares et al., 2013). Pupil diameter is relatively stable during alertness. This can be verified more clearly from the PSD plots for the same data as PSD represents the contribution of a particular frequency to the time series data.

PSD was computed using Welch periodogram method in Matlab®. As drowsiness is estimated to be present at higher frequency fluctuations, very low frequencies below are 0.03Hz are filtered using a Butterworth high pass filter. PSD of the filtered data points (Figure 11(b)) shows that P4 has higher amplitude fluctuations at relatively higher frequencies (until 0.4 Hz), followed by P3. It should be noted that task difficulty is high in P3 and P4 is a scenario with good visibility and benign weather conditions. Lack of alertness in P4 is hence due to pilot fatigue.

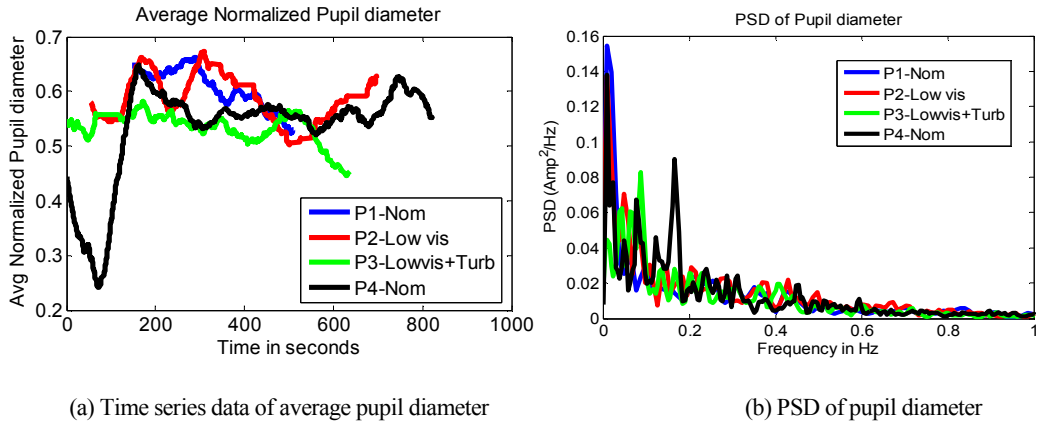

(a) Time series data of average pupil diameter
(b) PSD of pupil diameter

Figure 11. Pupil diameter plots

Now, the cognitive load experienced by the pilot in P3 and P4 can also be expressed through his control strategy. This exercise is done to establish a relationship between pilot's performance and his physiological measures, namely the pupil diameter. Figure 12 shows the PIW and power frequency plots of his commands on pitch and roll control columns. It can be seen that the pilot is most aggressive on his control columns in P4. In case of P3, he used more power in roll control column during 100-150 and 350-400 seconds. This was the phase where turbulence was introduced (Figure 13). Degraded visibility conditions did not cause much workload because pilot could still use his instruments for maintaining the flight path.

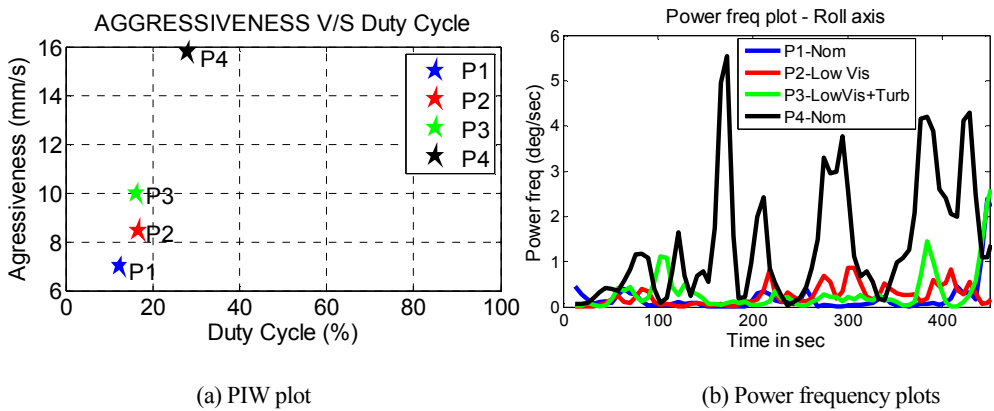

(a) PIW plot
(b) Power frequency plots

Figure 12. Pilot's control strategy on pitch and roll control columns

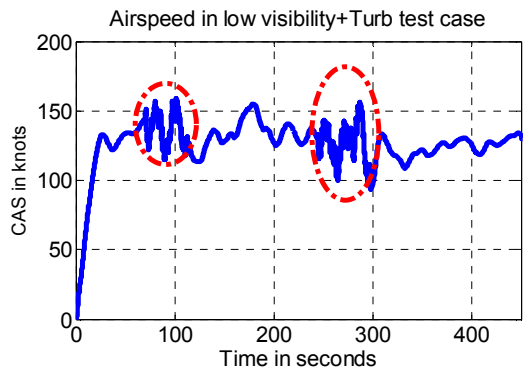

Figure 13. Airspeed in P3 test case

Hence, as per his control strategy measure also, P4 has caused the maximum workload, followed by P3. This corroborates with the eye parameter results.

## 3. Discussion

In the current study, we have discussed different potential applications of eye gaze measurement in an aircraft programme. The results are validated against standard statistical methods and pilot's subjective opinions.

Analysis results discussed in section 3.1 shows evidence of difference in pilot's visual scan behavior for different flight conditions. In nominal scenario, pilot tends to use more of the outside visual information to maintain flight path. However, his concentration is more on monitoring the concerned head down displays in the abnormal or emergency conditions. Overall results show that we can corroborate pilot's actions with his monitoring behavior. Hence the inference is that such a system can be used to design optimum flying task scenarios wherein pilots can make efficient searches of their displays to locate relevant information.

The results in this section also show that there is a significant difference in the durations for data processing for an experienced and a novice pilot. Novice pilots take longer time to view, process and decode information on the displays. However, more simulation exercises are required to make conclusions this regard because the effect sizes are very small.

Results from section 3.2 suggests the usage of ETS system to evaluate the effectiveness of a display interface. Pilot's performance in a flight segment is compared with his dwell time on each AOI to indicate the reason behind his good or poor performance. For instance, it was observed that airspeed control was poor for degraded visibility test case under first level segment. This was clearly related to insufficient monitoring of airspeed during the same period of time. Hence analyzing pilot's monitoring behavior in conjunction with simulation and event data can help in recreating the complete scenario with quantitative judgements. Another observation made in this section was that experienced pilots have better visual scan pattern than novice pilots. They monitor the most relevant displays at right times.

Furthermore, it is shown in section 3.3 that variations in pupil diameter can be an indication of pilot's workload levels. Cognitive workload experienced by the pilot under different task difficulties can be analyzed using PSD of pupil diameter.

Drowsiness is experienced when pilot is either bored or fatigued. Condition for fatigue is recreated in this study and results of PSD plots of pupil diameter shows that after long hours of simulator flying, participant's pupil size has more fluctuations and hence his alertness levels are low. These ocular measures are also compared with performance-based measures and both results are found to reflect the same conclusions.

## 4. Instructor's Dashboard for simulation and monitoring pilot's scan behaviour

Based on the understanding on the various applications of eye tracking, a support system is developed to assist the designers or evaluators in real time. The system is intended to be used in a flight simulator environment for design optimization of any display or display symbology. A snapshot of the user interface is shown in Figure 14. Real-time display of gaze fixation, scan path of gaze movement and scatterplot of the gaze location is provided to feedback on pilot's attention allocation. The application is a Visual C# based graphical user interface. The application receives eye gaze data at 60Hz over Ethernet using TCP/IP protocol. Live feed from the scene camera is also captured at 25 frames per second. The gaze location is overlaid on the video from the scene camera as shown in Figure 14.

The application also displays cognitive load parameters in real time as a 2-dimensional time series plot. The plots are updated every 1 second with the median value.

Using Eye Tracker to Evaluate Cockpit Design -- A Flight Simulation Study

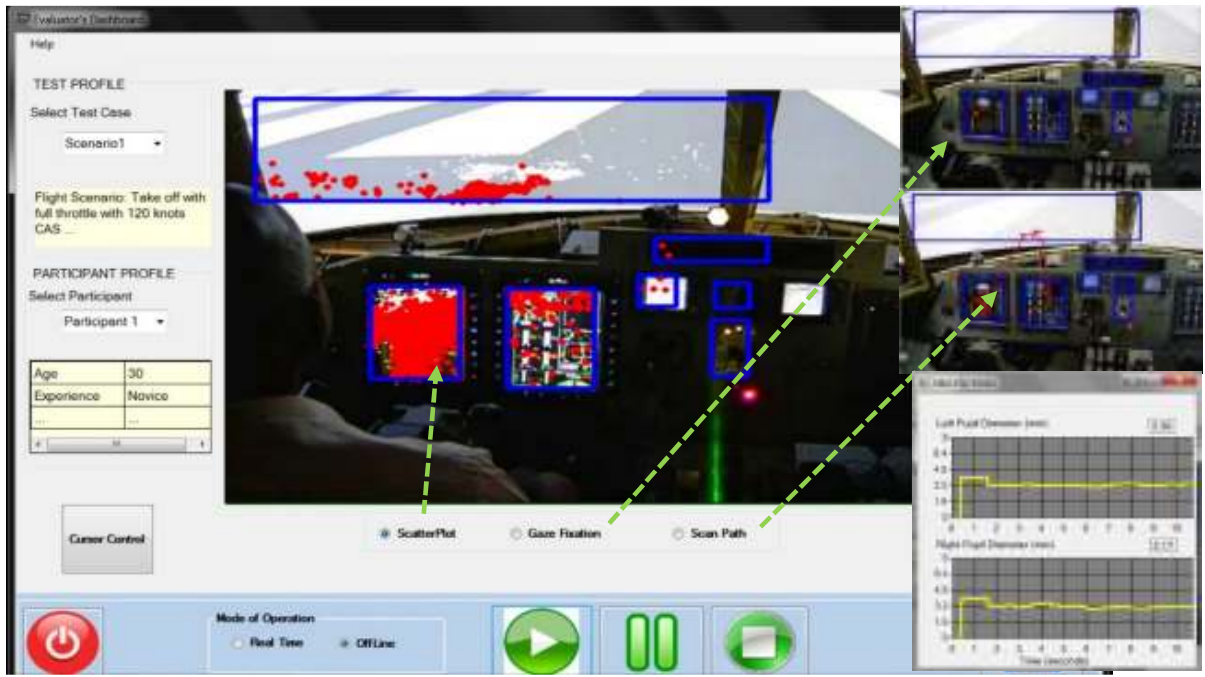

Figure 14. Evaluator's console

## Conclusions and Future work

Advances in technologies have allowed better measurement instruments. For example, measurement of brain or cardiac activity also gives good insight into mental workload. But these techniques remain intrusive and cumbersome. Eye tracking method used in this research work allows to measure pilot's reaction to test scenario without being intrusive. This provides a more natural environment for the pilots while conducting the experiments and therefore allows collection of more realistic data.

After a detailed literature survey, non-intrusive SmartEye based ETS system was selected, set up and tested at CSIR-NAL's flight training simulator. Pilot in the loop simulations were conducted for predefined test scenarios. The test scenario was a typical circuit and landing procedure to be followed by the pilot. Simulations were conducted for three different task scenarios: nominal, stall and the degraded visibility conditions. Selection of task scenarios were based on pilot's visual scanning requirements. While the exercise was proceeding, the ETS recorded information like scan path, scatter plot, mean fixation map, dwell time and fixation sequence. Analysis results are discussed with reference to areas of relevance. Inferences are made based on comparison between eye tracking data, pilot performance data and the flight parameters. The main findings from the study suggests that ETS is useful to decode pilot's monitoring behavior, to estimate pilot's cognitive load variations and to predict pilot's fatigue.

This exercise proves that the eye tracking system, in conjunction with performance metrics (based on deviation from the desired flight path data), can provide a significant insight into the design of aircraft from the perspective of human factors. To realize the said objective, an instructor or evaluator's console is developed that is discussed in section 5.

In a nutshell, initial analysis shows that use of eye gaze systems can help in the design of better aircraft systems and to develop better flying strategies. This shall ultimately help in reducing pilot's cognitive workload and improving situational awareness that shall enhance flight safety.

However, the basic limitation of this study is that the results presented herein are restricted to concept proving only and all the experiments are conducted by a single pilot. More simulation exercises need to be conducted in future with different mission scenarios and with more pilots to provide concrete evidences towards each of the application areas discussed in this paper.

## Acknowledgments

Authors sincerely thank the test pilot, Retd. Air Cmdr Jose Mathappan, for supporting this study with his valuable comments and support for data collection.

Using Eye Tracker to Evaluate Cockpit Design -- A Flight Simulation Study

## Author Contributions

Hebbar P. A. has contributed towards installation of ETS, formulation of the user study and conducting the experiments. Pashilkar A. A. and Biswas P. have contributed in overall guidance, discussion of the results and for reasoning and discussions.